\documentclass[12pt]{article}     
\newlength{\dinwidth}                       
\newlength{\dinmargin}                      
\def\lsim{\mathrel{\rlap{\lower4pt\hbox{\hskip1pt$\sim$}}
    \raise1pt\hbox{$<$}}}                
\def\gsim{\mathrel{\rlap{\lower4pt\hbox{\hskip1pt$\sim$}}
    \raise1pt\hbox{$>$}}}                
\input epsf
\epsfverbosetrue
\begin{document}
\begin{flushright}
GLAS--PPE/96--05\\
\today
\end{flushright}
\vspace*{.5cm}
\begin{center}  \begin{Large} \begin{bf}
Colour Coherence in Photon Induced Reactions \\
  \end{bf}  \end{Large}
  \vspace*{5mm}
  \begin{large}
A. Lebedev$^a$, L. Sinclair$^b$, E. Strickland$^b$, J. Vazdik$^a$.\\ 
  \end{large}
\end{center}
$^a$ P.N.Lebedev Physical Institute, Academy of Sciences of Russia,
     Leninsky Prospect~53, \\117924~Moscow, Russia\\
$^b$ Department of Physics and Astronomy, Glasgow University,
     G12~8QQ Glasgow, \\Scotland, U.K.\\
\begin{quotation}
\noindent
{\bf Abstract:}
Colour coherence in hard photoproduction is considered using the
Monte Carlo event generators PYTHIA and HERWIG.  Significant effects in
the parton shower are
found using multijet observables for direct and resolved 
photon induced reactions.
The particle flow in the interjet region of direct processes shows a
strong influence of string fragmentation effects.
\end{quotation}
\section{Introduction}
\begin{figure}[htb]
     \epsfxsize=7.cm
     \centering
     \leavevmode
     \epsfbox{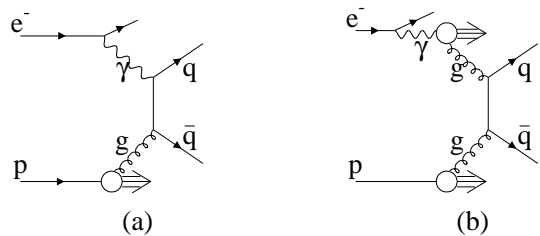}
     \caption{\label{f:gla1}\it Examples of (a) direct and (b) resolved 
             photoproduction}
     \end{figure}

        Colour coherence is an intrinsic property  of QCD.
     Its observation is important in the study of strong interactions
     and in the search for deviations from the Standard Model \cite{Dok}. It is 
     interesting to look for colour coherence effects in hard
     photoproduction processes at the $ep$-collider HERA
     where large momentum 
     transfers can be achieved and both direct 
     (Fig.~\ref{f:gla1}(a)) and resolved (Fig.~\ref{f:gla1}(b))
     photon induced events occur. 
     In Section~\ref{s:glasec} multijet observables are studied which reveal 
     coherence at the parton shower level for both direct and resolved 
     photoproduction.
     In Section~\ref{s:russec} consideration of the particle flow in direct 
     photoproduction shows colour coherence effects at the fragmentation stage 
     of hadron production.

\section{\label{s:glasec}Multijet observables}

     The effects of colour coherence on the emission pattern of jets in 
     $e^{+}e^{-}$ collisions are well known and intuitive.  However in 
     hadron--hadron collisions the large number of possible colour flows 
     involved in jet production complicates the identification of variables 
     sensitive to coherence.
     Here, radiation patterns in $\gamma p$ collisions are studied
     by considering events where soft radiation is hard enough to form a jet.
     This reduces the effect of secondary interactions in resolved 
     photoproduction.

     The effects arising from different implementations of coherence were
     studied using 500~pb$^{-1}$ of events generated with
     PYTHIA 5.7~\cite{Ben} and HERWIG 5.8~\cite{Hrw}.
     Direct and resolved events
     were generated separately and combined according to their cross sections.
     Events were generated with a minimum $p_T$ of 20~GeV using the GRV
     proton and photon structure functions~\cite{Glu}.
     Two event samples were generated using PYTHIA: the PYTHIA Coherent sample
     and the PYTHIA Incoherent sample which was obtained by switching off the 
     coherence in the parton shower and the initial-final state coherence.
     HERWIG represents an alternate implementation of coherent processes.
     Jets of particles were found using the KTCLUS~\cite{ktclus} 
     algorithm in covariant
     $p_T$ mode with radius equal to 1.
     Three jet events with
     at least two jets of transverse energy ($E_t^{jet}$) satisfying
     $E_t^{jet} > 30$~GeV
     and a third jet of $E_t^{jet} > 3$~GeV were selected.
     The jets are ordered by $E_t^{jet}$ decreasing and 
     referred to as ``first'', ``second'' and ``third'' jet acccordingly
     in the following.
     Two scenarios were considered, one to reflect the acceptance 
     in jet pseudorapidity ($\eta^{jet}$) of the
     present ZEUS detector, $|\eta^{jet}| < 2.5$, and one to show the
     possibilities with an extended acceptance, $|\eta^{jet}| < 4$.
     In addition the events satisfied $0.2 < y < 0.85$ and 
     $P^{2} < 4$~GeV$^{2}$, where $P^{2}$ is the negative of the 
     four-momentum squared of the photon. 

     An overall drop in cross section is observed between incoherent
     and coherent event samples. For example, with a luminosity of 
     250~pb$^{-1}$ and the standard detector acceptance, $|\eta^{jet}| < 2.5$,
     2600 multijet events are predicted by PYTHIA Incoherent, 
     1728 by PYTHIA Coherent and
     1665 events by HERWIG.  For comparison in the extended acceptance 
     scenario, $|\eta^{jet}| < 4$, 3012 multijet events are predicted 
     by HERWIG.

\begin{figure}[htb]
     \epsfxsize=16.cm
     \centering
     \leavevmode
     \epsfbox{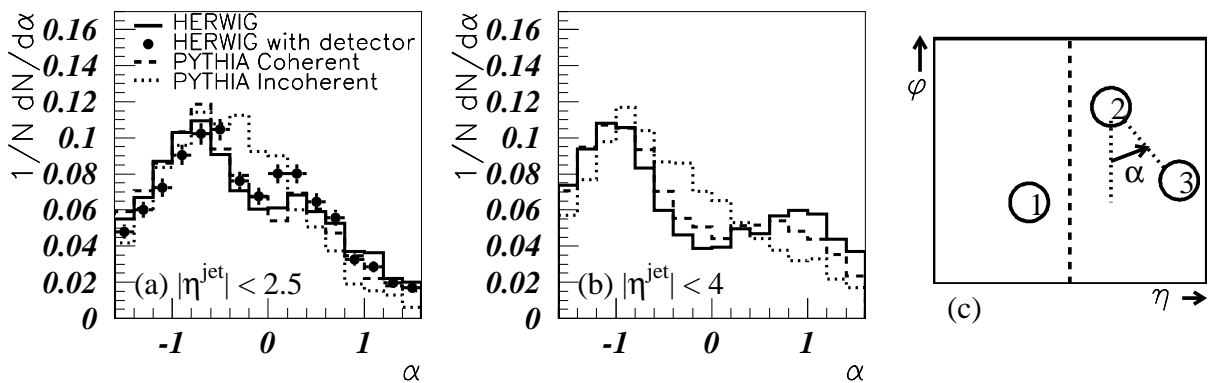}
     \caption{\label{f:gla2}\it $\alpha$ distributions (a) with the present 
     acceptance and (b) with extended acceptance.  (c) illustrates the 
     definition
     of $\alpha$ for a typical event geometry.  Jets are ordered
     by $E_t^{jet}$ decreasing.}
     \end{figure}

     The angular distribution of the third jet is also affected.  
     Following~\cite{CDF} the angle
     $\alpha$ is defined as the azimuthal angle of the third 
     jet about the second jet in the $\eta - \varphi$ plane.
     Here, however, we use centre-of-mass (c.m.) variables so 
     $\alpha = \arctan (\Delta H / |\Delta \varphi|)$ where 
     $\Delta H = \mbox{sign}(\eta_2^{cm}) (\eta_3 - \eta_2)$ and
     $\eta_2^{cm} = \eta_2 - 1/2 (\eta_2 - \eta_1)$ 
     and $\Delta \varphi = \varphi_3 - \varphi_2$.
     $\eta_1$, $\eta_2$ and $\eta_3$ refer to the pseudorapidities in the
     lab frame of the first, second and third jets
     respectively and positive $\eta$ is in the direction 
     of the incoming proton. 
     $\Delta \varphi$ is the difference in azimuth ($\varphi$) between the 
     second and third jets (in radians).
     The definition of $\alpha$ is illustrated for
     a typical event geometry in 
     Fig.~\ref{f:gla2}(c).
     The distribution of $\alpha$ as shown in Fig.~\ref{f:gla2}(a) 
     is broader for coherent events.  This is consistent with our understanding 
     that for
     coherent processes radiation is generally suppressed in regions far from
     the directions of the incoming coloured partons.
     In addition, reducing the bias on the distribution by 
     increasing the acceptance
     from $|\eta^{jet}| < 2.5$ to $|\eta^{jet}| < 4$ produces a more
     pronounced depletion in
     the central region for coherent events (Fig.~\ref{f:gla2}(b)).

     Canonical detector effects were simulated by smearing the
     HERWIG jet quantities with  Gaussian distributions of varying widths. A 
     resolution of 20\% (10\%) was used to smear the $E_t^{jet}$ of jets 
     with $E_t^{jet} < 10$~GeV ($E_t^{jet} \ge 10$~GeV).
     The width of the difference between generated and 
     detected values of $\eta^{jet}$ and $\varphi^{jet}$ was taken to be 0.1. 
     As shown in Fig.~\ref{f:gla2}(a) such detector effects should not seriously 
     hinder the measurement of $\alpha$ distributions.
      
\begin{figure}[htb]
     \epsfxsize=16.cm
     \centering
     \leavevmode
     \epsfbox{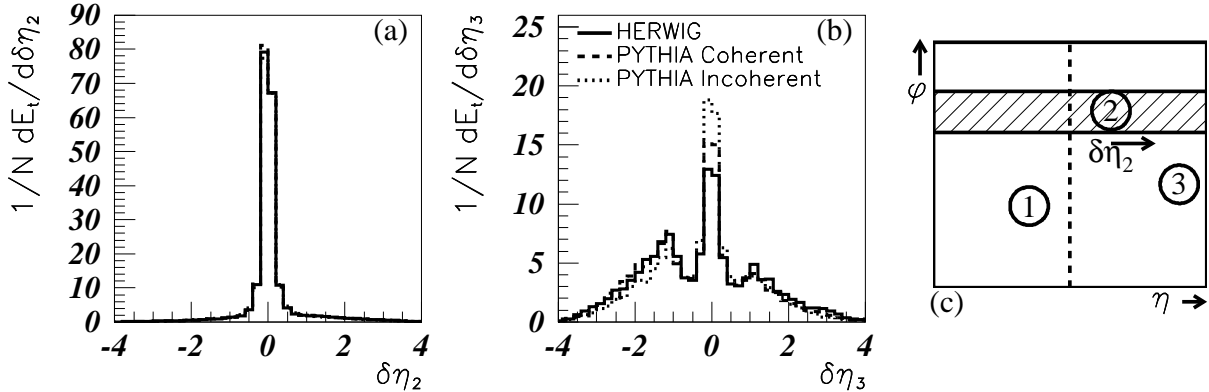}
     \caption{\label{f:gla3}\it Jet profiles for (a) the second 
     and (b) the third highest $E_t^{jet}$ jets for the extended acceptance
     scenario.  The definition of the jet profile is illustrated in (c) for
     the second jet.}
     \end{figure}

     The coherent emission of soft radiation does not have a strong
     effect on the jet profiles of the first and second jets.
     For instance, in Fig.~\ref{f:gla3}(a) the transverse energy profile of the 
     second jet is shown.
     This is the distribution of
     $\delta\eta_2 = \eta^{part} - \eta_2$,
     where $\eta^{part}$ is the $\eta$ of a 
     particle within one radian of $\varphi$ of the jet centre, weighted 
     by the transverse energy of the particle, as illustrated in 
     Fig.~\ref{f:gla3}(c).  For this the extended acceptance scenario particles
     are considered with absolute $\eta$ up to 5.  
     The profile of the third jet is shown in 
     Fig.~\ref{f:gla3}(b).  The occurrence of two peaks outside the
     jet core is due to partial overlap in $\varphi$ of the first 
     or second jet.
     A strong effect of coherence is apparent; it 
     leads to less energy in the core of the third jet.

\begin{figure}[htb]
     \epsfxsize=16.cm
     \centering
     \leavevmode
     \epsfbox{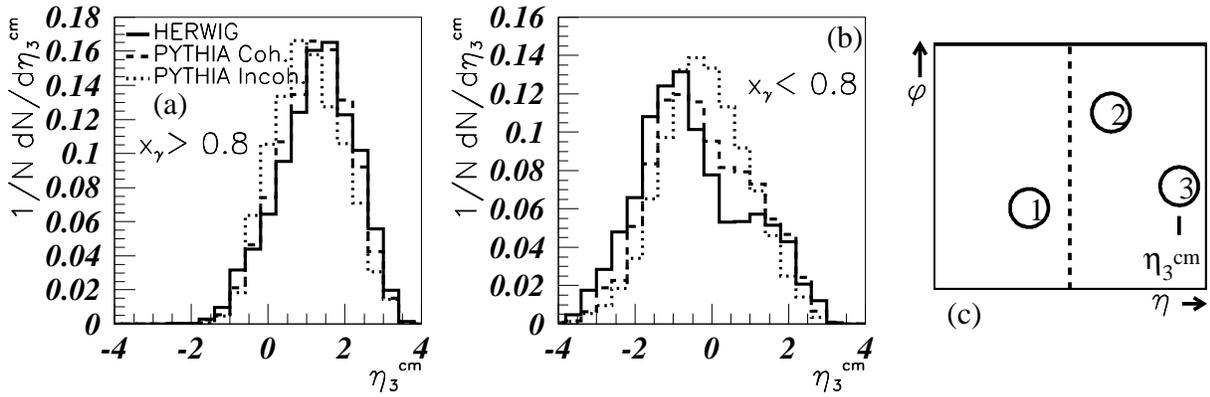}
     \caption{\label{f:gla4}\it $\eta_3^{cm}$ distributions 
     separated by an $x_{\gamma}$ cut of $0.8$ into (a) direct and (b) resolved
     samples for the extended acceptance scenario of $|\eta^{jet}| < 4$.
     In (c) the definition of $\eta_3^{cm}$ is illustrated.}
     \end{figure}

     One of the anticipated effects of colour coherence is that radiation from
     an incoming parton should be inhibited in regions far from the
     initial partons direction.
     Therefore in  direct photoproduction events, where the single coloured 
     parton in the initial state has positive $\eta$, the coherent 
     emission pattern should be at relatively higher $\eta$ than the 
     incoherent emission. 
     We have selected a subsample of events which is 
     enriched in direct photon events 
     by requiring $x_{\gamma} > 0.8$ where 
     $x_{\gamma} = (\sum_{jets} E_t^{jet} e^{-\eta^{jet}}) / 
                            (2 E_{\gamma})$.
     The sum runs only over the two highest $E_T^{jet}$ jets and
     $E_{\gamma}$ is the energy of the incoming photon.
     The $\eta$ of the third jet
     in the c.m. frame,
     $\eta_3^{cm} = \eta_3 - 1/2 (\eta_2 - \eta_1)$, is shown
     for this selection in Fig.~\ref{f:gla4}(a).
     The expected effective enhancement of radiation at large $\eta_3^{cm}$
     can clearly be seen.

     Resolved events, with incoming coloured partons from both the $\gamma$
     and $p$ directions, should show an effective enhancement of 
     radiation in coherent processes
     both at higher positive and at more negative pseudorapidities in 
     comparison to incoherent emission.  
     This effect is evident as shown in Fig.~\ref{f:gla4}(b).  Note that the 
     extended acceptance scenario must be employed in order to see the 
     relative enhancement of radiation at high $\eta$ in resolved events.

     To summarize this section, a high integrated luminosity 
     ($\sim 250$pb$^{-1}$) is desirable in order
     to accumulate statistics in multijet events at high $E_t^{jet}$.  However
     luminosity upgrades which involve a significant reduction of forward 
     acceptance are not worthwhile for this study.
     They destroy the sensitivity to colour coherence without significantly
     improving the statistical uncertainty.

\section{\label{s:russec}Interjet string effects in direct photoprocesses}
\begin{figure}[h!]
   \epsfxsize=15cm
   \centering
   \leavevmode
   \epsfbox{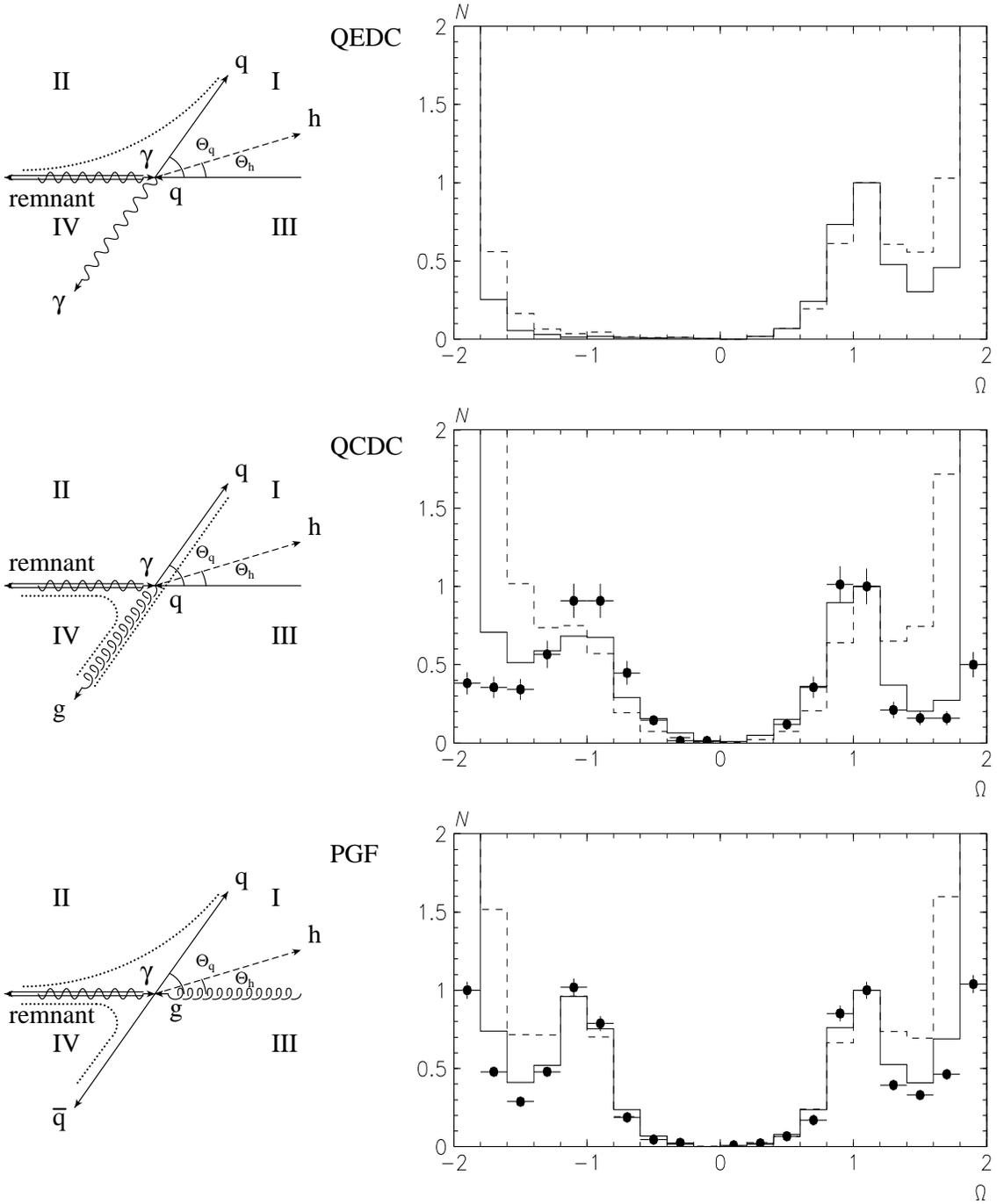}
   \caption[junk]{{\label{f:rus1}\it 
  Left: Topology of direct photoproduction processes 
      in  the $\gamma-parton$ c.m. frame.
 The double line is the proton remnant, dotted - strings.
                $\theta_q$ and $\theta_h$ are
            quark and hadron emission angles.
  Right:  Charged particle flow $N$ 
          normalized to 1 at the quark emission angle
          vs the scale angle $\Omega$.
   Solid line - generator level SF, dashed line -  generator level IF,
   dark circle - H1 simulation SF.
   In the regions I - IV $\Omega$ is changing within the limits: 
   $0\leq\Omega\leq1$ (I), $1\leq\Omega\leq2$ (II),
   $-1\leq\Omega\leq0$ (III), $-2\leq\Omega\leq-1$ (IV).
     }}
     \end{figure}

        Colour coherence effects which lead to a change in particle
     flow $N$ distributions in the interjet region should be rather
     pronounced in the direct photon induced processes such as
     QED Compton on quark (QEDC), QCD Compton (QCDC) and Photon
     Gluon Fusion (PGF). 
       These distributions are considered here using the
     PYTHIA~\cite{Ben} generator with string (SF) or
     independent parton fragmentation (IF) into
     hadrons. Using SF is equivalent to taking into account the coherence 
     effects at the hadronization phase of event generation. The flow $N$ 
     depends on the string topology 
     and colour antennae which are different for the three 
     direct processes as shown in Fig.~\ref{f:rus1}(left).  

         The calculation at the generator level was done using the GRV proton 
     structure function~\cite{Glu} and minimum $p_T$ equal to 2.0~GeV.
     The H1 detector simulation was taken
     into account as well.   
       A jet-cone algorithm~\cite{cone} with radius equal to 1 was used for 
     the selection of two jets and gamma-jet events with 
     $E_t^{jet \, \rm or \, \mit \gamma}>3$~GeV and 
     jet (or final $\gamma$) emission angles $25-155^o$.  This procedure
     corresponds to the selection of mainly direct
     processes. 
        The calculated flow of charged particles with $p_t>0.2$~GeV
     emitted at angles of less than $20^o$ to the reaction plane
     is shown in Fig.~\ref{f:rus1}(right) as a function of the scale angle 
     $\Omega$.
     $\Omega$ is defined~\cite{Kho} as the ratio of the particle angle
     $\theta_h$ to the angle between
     partons.  $\Omega=0$ corresponds to the direction of the initial state 
     photon;
     $\Omega=1$ -- the final state quark;
     $\Omega=-1$ -- the final state
     $\gamma$, gluon or antiquark for QEDC, QCDC or PGF respectively;
     $\Omega=-2$ or 2 -- the proton remnant.
 
     It is seen that in the scale angle region between 1 and 2
     (region II in Fig.~\ref{f:rus1})  
     SF (solid histogram, generator level) and 
     IF (dashed histogram, generator level) 
     give different predictions for $N$. SF taking
     into account colour forces leads to a suppression of particle flow
     which is especially strong for QCDC process.
     The H1 detector simulation (dark circles, SF)
     weakly distorts the generator level $N$ distribution except for
     directions close to remnant proton emission where detector 
     acceptance is rather low. Thus colour coherence effects can be observed
     at the detector level.
     
\begin{figure}[htb]
   \epsfxsize=13.cm
   \centering
   \leavevmode
   \epsfbox{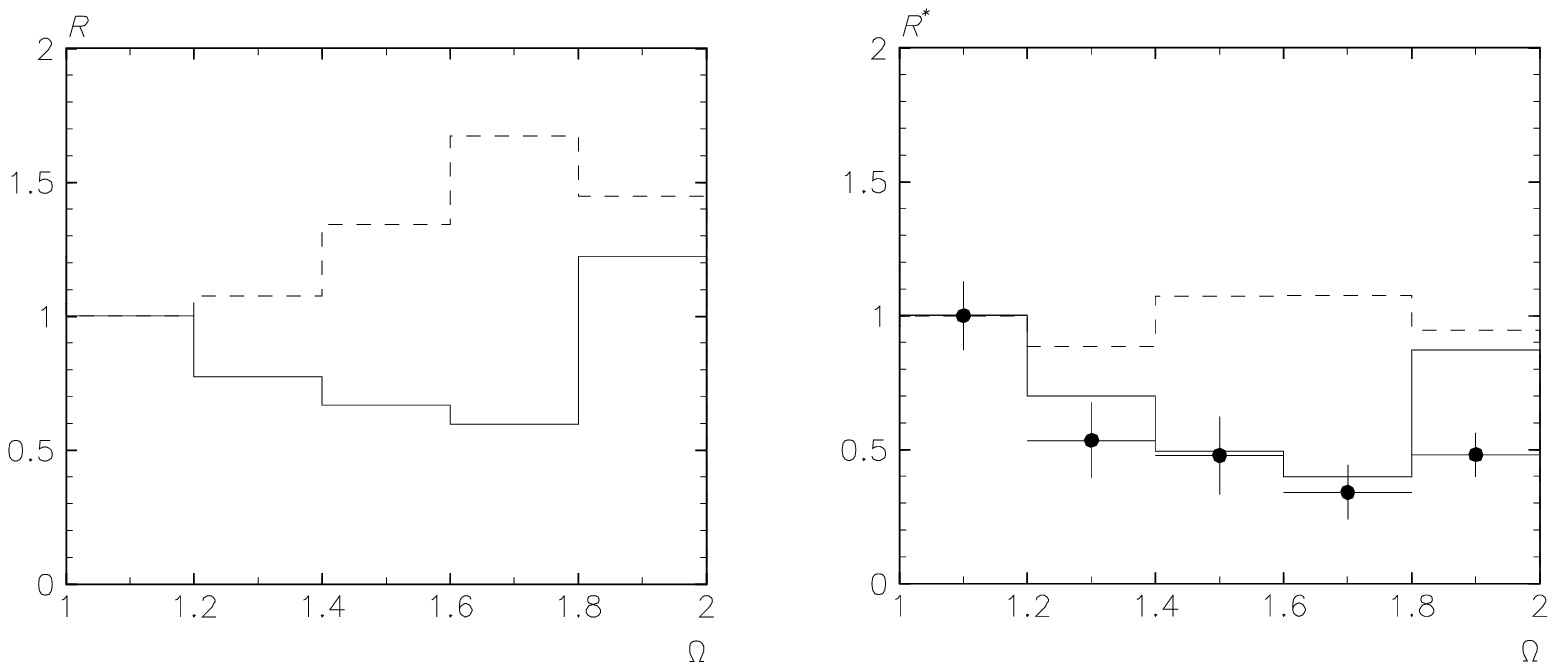}
   \caption[junk]{{\label{f:rus2}\it 
    Ratios $R$ and $R^{*}$ in the scale angle region
    $1<\Omega<2$. Notations are as in Fig.~\ref{f:rus1}.
     }}
     \end{figure}

        It is interesting to consider ratios of particle flows $N$
     for different processes since the ratio is less sensitive 
     to experimental errors.     
       The ratios 
    $R=N(QCDC)/N(QEDC)$, $R^{*}=N(QCDC)/N(PGF)$
     are shown in Fig.~\ref{f:rus2}. 
      Fig.~\ref{f:rus2} displays more clearly the role of 
     colour coherence which leads to drag effects in particle
     distributions.
 	It is seen that for the case of SF
     the suppression in $N$ for QCDC is more
     pronounced than for QEDC and PGF. At $\Omega = 1.7$ the
     suppression reaches a factor of  $\sim 3$.
       It has been found that misidentification of quark 
     and gluon jets for QCDC does not change this conclusion.

       To observe colour coherence at the fragmentation stage of
     hadron production in direct processes it is necessary
     to distinguish these processes from resolved photoproduction and
     to separate QEDC, QCDC and PGF from each other.
       The jet selection procedure used here enriches the data sample
     with direct processes. Further enrichment can be achieved
     by going to larger $E_t^{jet \, \rm or \, \mit \gamma}$ and by choosing 
     events with $x_{\gamma}$ close to 1. 
        Since the direct processes cross section is low we expect
        $\sim$~230 QEDC events at the detector level for an integrated
        luminosity of $\sim$~100~pb$^{-1}$.  So higher luminosity is
        needed to study interjet coherence.
   
\section{Conclusions}
        The observation of colour coherence in photoproduction
        processes is an important challenge, particularly given the
        unique opportunity at HERA to study direct as well
        as resolved photon induced reactions.
        Since the cross sections of multijet events or of prompt
        photon reactions are small high luminosity
        $ep$-collisions are necessary for their study.  
        250~pb$^{-1}$ would appear to be barely sufficient for these studies; 
        however in upgrading to 1000~pb$^{-1}$ it is essential that the forward
        acceptance should not be reduced.

\end{document}